\documentstyle[12pt]{article}
\makeatletter
\newcommand{\beq}{\begin{equation}}
\newcommand{\eeq}{\end{equation}}
\newcommand{\beqy}{\begin{eqnarray}}
\newcommand{\eeqy}{\end{eqnarray}}

\begin{document}
\begin{flushleft}
{\it  Yukawa Institute Kyoto}
\end{flushleft}
\begin{flushright}
 YITP-97-22\\
 hep-th/9705002\\
 May 1997
\end{flushright}
\renewcommand{\thefootnote}{\fnsymbol{footnote}}
\vspace{0.5in}
\begin{center}\Large{\bf 
The Relation between Mixed Boundary States\\ 
in Two- and Three- Matrix Models
}\\
\vspace{1cm}
\normalsize\ Masahiro Anazawa\footnote[1]
{Supported by JSPS. E-mail address: anazawa@yukawa.kyoto-u.ac.jp .},
Atushi Ishikawa\footnote[2]{
Supported by JSPS. E-mail address: ishikawa@yukawa.kyoto-u.ac.jp .}
$\quad$and$\quad$ Hirokazu Tanaka\footnote[3]{E-mail address:
hirokazu@yukawa.kyoto-u.ac.jp .}
\vspace{0.5in}

        Yukawa Institute for Theoretical Physics \\
        Kyoto University, Kyoto 606-01, Japan\\
\vspace{0.1in}
\end{center}
\renewcommand{\thefootnote}{\arabic{footnote}}
\setcounter{footnote}{0}
\vspace{0.4in}
\begin{abstract}
We discuss the relation among some disk amplitudes with
non-trivial boundary conditions in two-dimensional quantum gravity.
They are obtained by the two-matrix model as well as the 
three-matirx model
for the case of the
tricritical Ising model.
We examine them for simple spin configurations,
and find that a finite number of 
insertions of the different spin states
cannot be observed in the continumm limit.
We also find a closed set of eight Schwinger-Dyson equations which
determines the disk amplitudes in the three-matrix model.

\end{abstract}


\newpage

\baselineskip 17pt


It is well known that
two-dimensional gravity can be described by the matrix models. 
In fact, the $(p, q)=(2, 2m-1)$ models coupled to 2d gravity
can be realized by 
one-matrix models at the $m$-th critical points \cite{KD, K}.
It was conjectured that 
the general $(p, q)$ minimal models coupled to 2d gravity can be also 
described
by the two-matirx model,
if we consider the critical behaviors near $(p, q)$ multi-critical
points \cite{KBK, Douglas}.
This was explicitly shown in \cite{Tada, DKK}.
The two-matrix model, however, does not have the degrees of freedom
which represent the general boundary states of the matter configurations.
In order to examine such general boundary states,
 we should consider the multi-matrix models \cite{HI}.
In this paper, 
we treat the tricritical Ising model 
( $(p, q) = (4, 5)$ ),
which can be described by the two-matrix model as well as
the three-matirx model.
We concentrate our attention on the disk amplitudes and
find a relation between them calculated in
both matrix models.


Let us begin with the two-matrix model.
As an action which realizes the $(4, 5)$ model, we take the $Z_2$ 
symmetric ($A\leftrightarrow B$) one,
\begin{eqnarray}
S(A,B) &=& \frac{N}{\Lambda} tr \left\{ U(A) + U(B) - AB \right\},
\label{eq:Action}\\
U(\phi) &=& 8{\phi}+2{\phi}^2+\frac{8}{3}{\phi}^3+\frac{1}{4}{\phi}^4.
\nonumber 
\end{eqnarray}
Here $\Lambda$ denotes a bare cosmological constant.
The critical potential
$U(\phi)$ 
can be determined by
the orthogonal polynomial method \cite{DKK}.
We can easily find that the following four Schwinger-Dyson
equations close \cite{GN, Staudacher}:
\begin{eqnarray}
0 & = & \int [dAdB] tr 
        \left( 
        \frac{\partial}{\partial A} A^{n-1} B^{k}
        \right)
          e^{-S(A,B)} ; ~~k = (0, 1, 2) ,
\nonumber \\
0 & = & \int [dAdB] tr 
        \left( 
        \frac{\partial}{\partial B} A^{n} B 
        \right)
          e^{-S(A,B)}.
\label{eq:S-D eq}
\end{eqnarray}
Introducing a notation 
$w^{(k)}_n=\frac{\Lambda}{N}\langle tr (A^n B^k) \rangle$,
we can write them as
\begin{eqnarray}
0 & = & \sum_{l=0}^{n} w^{(0)}_l w^{(k)}_{n-l}
        - 8w^{(k)}_{n+1}-4w^{(k)}_{n+2}-8w^{(k)}_{n+3}
        -w^{(k)}_{n+4}+w^{(k+1)}_{n+1},
\nonumber \\
0 & = & 8w^{(0)}_n + 4w^{(1)}_n +8w^{(2)}_n + w^{(3)}_n - w^{(0)}_{n+1},
\label{eq:S-D eq2}
\end{eqnarray}
in the large $N$ limit. 
It is convenient to use the resolvent representation
$W^{(k)}(p) = \sum^{\infty}_n w^{(k)}_n p^{-(n+1)}$. From
eqs.(\ref{eq:S-D eq2}), we obtain
\begin{eqnarray}
0 & = & [W^{(0)}(p)-v(p)]W^{(k)}(p)+W^{(k+1)}(p)+a^{(k)}(p),
\nonumber \\ 
0 & = & (8-p)W^{(0)}(p)+4W^{(1)}(p)+8W^{(2)}(p)+W^{(3)}(p)+\Lambda,
\label{eq:S-D eq3}
\end{eqnarray}
where
\begin{eqnarray}
v(p) &=& 8+4p+8p^2+p^3,
\nonumber \\
a^{(k)}(p) &=& (4+8p+p^2)w^{(k)}_0 + (8+p)w^{(k)}_1 + w^{(k)}_2.
\end{eqnarray}
Note that $w^{(0)}_0=\Lambda$
and we used the $Z_2$ symmery. 
We can eliminate $W^{(1)}(p)$, $W^{(2)}(p)$ and $W^{(3)}(p)$, 
and have a fourth order equation of $W^{(0)}(p)$:
\begin{equation}
V(p)^4+\alpha_3(p) V(p)^3+\alpha_2(p) V(p)^2+\alpha_1(p) V(p)
+\alpha_0(p) =0,
\label{45disk1}
\end{equation}
where
\begin{eqnarray}
V(p) & = & W^{(0)}(p) - v(p)
\nonumber \\
\alpha_3(p) & = & v(p)-8,
\nonumber \\
\alpha_2(p) & = & 4 - 8v(p) + a^{(0)}(p),
\\
\alpha_1(p) & = & p - 8 + 4v(p) -8a^{(0)}(p) - a^{(1)}(p),
\nonumber \\
\alpha_0(p) & = & -\Lambda +(p-8)v(p) + 4a^{(0)}(p) + 8a^{(1)}(p) +
                  a^{(2)}(p).
\nonumber 
\end{eqnarray} 
We must provide the amplitudes $w^{(0)}_1$, $w^{(0)}_2$, $w^{(1)}_1$,
$w^{(1)}_2$ and $w^{(2)}_2$ in order to get $V(p)$.
These can be determined by the orthogonal polynomial method
\cite{DKK} with tedious calculations.
Since fourth order equations 
can be solved in general,
we can obtain $V(p)$ explicitly.
On the other hand, in the case of the three-matrix model, the corresponding
Schwinger-Dyson equation will be found to be of sixth order.
We take, therefore, another method to examine $V(p)$.

The continuum limit can be carried out by the renormalizaion;
$\Lambda = 70(1-a^2 t)$ and $p = a \zeta$
with $a$ the lattice spacing ($a \rightarrow 0$) \cite{GM}. Here 
$70$ is a critical value of $\Lambda$, and
($t$, $\zeta$) are the renormalized bulk and boundary
cosmological constant respectively.
Let us denote the continuum universal part of $W^{(k)}(p)$
as $w^{(k)}(\zeta, t)$.
We know the scaling behavior of it as
\begin{eqnarray}
V(p) = X + Y \zeta a + w^{(0)}(\zeta, t) a^{5/4} + {\cal O} (a^{6/4}),
\label{eq:scaling}
\end{eqnarray} 
where $X$ and $Y$ are some constants.
The first and second terms are non-universal parts.
Substituting eq.(\ref{eq:scaling}) into 
eq.(\ref{45disk1}) and expanding it in $a$,
we can obtain an equation for each order of $a$.
We find that $(X, Y) = (0, -1)$
and that the term 
${\cal O} (a^{6/4})$ does not contribute to the decision of 
$w^{(0)}(\zeta, t)$.
At the same time,
we obtain a relation which is satisfied by
$w^{(0)}(\zeta, t)$:
\begin{eqnarray}
w^{(0)}(\zeta, t)^4 -4 t^{5/4} w^{(0)}(\zeta, t)^2 
+ 2(t^{5/2}-5 t^2 \zeta +20 t \zeta^3 - 16 \zeta^5) = 0,
\end{eqnarray} 
after appropriate rescalings of $t$ and $\zeta$.
We can then identify the continuum universal disk amplitude as
\begin{eqnarray}
w^{(0)}(\zeta, t) = \left(\zeta + \sqrt{\zeta^2 - t} \right)^{5/4} 
              + \left(\zeta - \sqrt{\zeta^2 - t} \right)^{5/4}.
\end{eqnarray} 

Our aim is to compare the disk amplitudes which have the mixed boundary
states.
These $w^{(k)}(\zeta, t)$
can be easily calculated with the use of the relations between 
$W^{(k)}(p)$, eqs.(\ref{eq:S-D eq3}).
We can identify
\begin{eqnarray}
w^{(k)}(\zeta, t) & = & (-1)^{k}
\left[
                               w^{(0)}(\zeta, t)
\right]^{k+1}.
\label{result-2}
\end{eqnarray}  
The insertion of the matrix $B$ on the boundary 
results in the multiplication of
the amplitude $ w^{(0)}(\zeta,t) $.
Here we identified the universal and non-universal parts as follows.
If there are polynomials of $\zeta$ {\it or} $t$, they are 
regarded as non-universal.
And, if there are amplitudes constituted by the product of a 
universal amplitude and a non-universal part, they are non-universal.
This rule is a little different from the one used in 
\cite{IIKMNS}.
For example, in the case of the Ising model $(3, 4)$,
there is a $t^{4/3}$ term in the corresponding $W^{(1)}(p)$,
the scaling behavior of which 
is the same as the one of $w^{(1)}(\zeta, t)$; $a^{8/3}$.
The coefficient of this term is $4$, if we use the third order potential.
It changes, however, to $8/3$ when we take the fourth order potential.
This shows us that the $t^{4/3}$ term is non-universal.
In our rule, 
we thus should identify polynomials of $\zeta$ {\it or} $t$
as the non-universal parts.


Next, let us consider the three-matrix model.
We take the $Z_2$ symmetric 
$(A \leftrightarrow C)$
action
\begin{eqnarray}
S(A,B,C) & = & \frac{N}{\Lambda} tr
\left\{ U_1(A) + U_2(B) + U_1(C) - A B - B C
\right\},
\nonumber \\
U_1(\phi) & = & - \frac{111}{16} \phi + \frac{9}{4} \phi^2
                + \frac{1}{3} \phi^3,
\\
U_2(\phi) & = & \frac{3}{4} \phi^2 - \frac{1}{12} \phi^3. 
\nonumber
\label{Action2}
\end{eqnarray}
We find that the following eight Schwinger-Dyson equations close:
\begin{eqnarray}
0 & = & \int [dAdBdC] 
tr \left( \frac{\partial}{\partial A} A^{n-1} B^k \right)
e^{-S(A,B,C)}, 
\nonumber \\
0 & = & \int [dAdBdC] 
tr \left( \frac{\partial}{\partial A} BCA^{n-1} \right)
e^{-S(A,B,C)}, 
\nonumber \\
0 & = & \int [dAdBdC] 
tr \left( \frac{\partial}{\partial A} C^{k+1} A^{n-1} \right)
e^{-S(A,B,C)};~~~k=(0, 1),
\label{eq8}\\
0 & = & \int [dAdBdC] 
tr \left( \frac{\partial}{\partial B} C^k A^{n} \right)
e^{-S(A,B,C)},
\nonumber \\
0 & = & \int [dAdBdC] 
tr \left( \frac{\partial}{\partial C} A^{n} B \right)
e^{-S(A,B,C)}.
\nonumber
\end{eqnarray}
Introducing the notations
$w^{(k,l)}_n = \frac{\Lambda}{N}\langle tr (A^n B^k C^l) \rangle$
and
$W^{(k,l)}(p) = \sum^{\infty}_n w^{(k,l)}_n p^{-(n+1)}$,
we may write them in the resolvent
representation,
\begin{eqnarray}
0 & = & [W^{(0, 0)}(p)-u(p)]W^{(k, 0)}(p) 
      + W^{(k+1,0)}(p) + a^{(k, 0)}(p), 
\nonumber \\
0 & = & [W^{(0, 0)}(p)-u(p)]W^{(1,1)}(p)
      + W^{(2,1)}(p) + a^{(1,1)}(p), 
\nonumber \\
0 & = & [W^{(0, 0)}(p)-u(p)]W^{(0,k+1)}(p)
      + W^{(1,k+1)}(p)+a^{(0,k+1)}(p), 
\label{eq: S-D 45}\\
0 & = & \frac{3}{2} W^{(1,k)}(p) - \frac{1}{4} W^{(2,k)}(p)
        - W^{(0,k+1)}(p) - p W^{(0,k)}(p) + w^{(0,0)}_k,
\nonumber \\
0 & = & -\frac{111}{16}W^{(1,0)}(p) + \frac{9}{2} W^{(1,1)}(p)
        + W^{(1, 2)}(p) - W^{(2, 0)}(p),
\nonumber 
\end{eqnarray}
where
\begin{eqnarray}
u(p) & = & -\frac{111}{16} + \frac{9}{2} p + p^2,
\nonumber \\
a^{(k,l)}(p) & = & (\frac{9}{2} + p) w^{(k, l)}_0 + w^{(k, l)}_1. 
\nonumber 
\nonumber
\end{eqnarray}
Here $w^{(0, 0)}_0=\Lambda$ and we used the $Z_2$ symmetry.
By eliminating
$W^{(1,0)}(p)$, $W^{(0,1)}(p)$, $W^{(2,0)}(p)$, $W^{(1,1)}(p)$, 
$W^{(0,2)}(p)$,
$W^{(2,1)}(p)$ and $W^{(1,2)}(p)$
from eq.(\ref{eq: S-D 45}), we obtain the following sixth order 
equation of
$W^{(0, 0)}(p)$,
\begin{equation}
U (p)^6+ \beta_5(p) U (p)^5 + \beta_4(p) U(p)^4 
+ \beta_3(p) U(p)^3 + \beta_2(p) U(p)^2 
+ \beta_1(p) U(p) + \beta_0(p) 
 = 0.
\end{equation}
Here
\begin{eqnarray}
U(p) & = & W^{(0, 0)}(p) - u(p),
\nonumber \\
\beta_5(p) & = & 12 + u(p), 
\nonumber \\
\beta_4(p) & = & 18 + 8p + a^{(0,0)}(p) +12u(p),
\nonumber \\
\beta_3(p) & = & - 92 +48p + 12a^{(0,0)}(p) - a^{(1,0)}(p)
                  -4\Lambda +(18+8p)u(p),
\nonumber \\
\beta_2(p) & = & -111 - 72p + 16p^2 + (18+4p)a^{(0,0)}(p)
                  -6a^{(1,0)}(p) - 4a^{(0,1)}(p) 
\nonumber \\
                && -24\Lambda + (-92+48p)u(p),
\\
\beta_1(p) & = & (-92+24p)a^{(0,0)}(p) + (18-4p)a^{(1,0)}(p)
                  - 24a^{(0,1)}(p) + 4a^{(1,1)}(p)
\nonumber \\
                && + (72-16p)\Lambda +16w^{(0,0)}_1 + (-111-72p+16p^2)u(p),
\nonumber \\
\beta_0(p) & = & -111a^{(0,0)}(p) - 16a^{(1,0)}(p) + 72a^{(0,1)}(p) 
                  + 16a^{(0,2)}(p).
\nonumber
\end{eqnarray}

We may take the same method used in the case of the two-matrix model.
The amplitudes
$w^{(0,0)}_1$, $w^{(0,0)}_2$, $w^{(1,0)}_0$, $w^{(1,0)}_1$,
$w^{(0,1)}_1$, $w^{(1,1)}_1$ and $w^{(0,1)}_2$
are determined by the orthogonal polynomial method \cite{AIT}.
With the renormalization; 
$\Lambda = -35(1-a^2 t)$ and $p = a \zeta$,
the scaling behavior for $U(p)$ is set as
\begin{eqnarray}
U(p) = X' + Y' \zeta a + w^{(0,0)}(\zeta, t) a^{5/4} + {\cal O} (a^{6/4}).
\end{eqnarray}
We find that $(X', Y') = (0, -\frac{4}{3})$ 
and identify
\begin{eqnarray}
w^{(0,0)}(\zeta, t) = \left(\zeta + \sqrt{\zeta^2 - t} \right)^{5/4} + 
            \left(\zeta - \sqrt{\zeta^2 - t} \right)^{5/4},
\end{eqnarray}
from the equation with $a^6$ order, after appropriate rescalings of 
$t$ and $\zeta$.
As expected, this coincides with the result for 
$w^{(0)}(\zeta, t)$ in the two-matrix model.

The disk amplitudes which have  mixed boundary states
can also be calculated by using the relations between 
$W^{(k,l)}(p)$, eqs.(\ref{eq: S-D 45}).
The results are
\begin{eqnarray}
w^{(1,0)}(\zeta, t) &=& -\left[
                   w^{(0,0)}(\zeta, t)^2 
                   \right],
\nonumber \\
w^{(2,0)}(\zeta, t) &=& -w^{(0,0)}(\zeta, t) w^{(1,0)}(\zeta, t),
\nonumber \\
w^{(0,1)}(\zeta, t) &=& \frac{3}{2} w^{(1,0)},
\nonumber \\
w^{(1,1)}(\zeta, t) &=& -w^{(0,0)}(\zeta, t) w^{(0,1)}(\zeta, t),
\label{result-3} \\
w^{(2,1)}(\zeta, t) &=& -w^{(0,0)}(\zeta, t) w^{(1,1)}(\zeta, t),
\nonumber \\
w^{(0,2)}(\zeta, t) &=& \frac{3}{2} w^{(1,1)}(\zeta, t),
\nonumber \\
w^{(1,2)}(\zeta, t) &=& -w^{(0,0)}(\zeta, t) w^{(0,2)}(\zeta,t).
\nonumber
\end{eqnarray}
Here we take the same rule as the one used
in identifying the universal and 
non-universal parts.


Now, let us compare the disk amplitudes calculated in the two- and
three-matrix models. From the results eqs.(\ref{result-2}) 
and eqs.(\ref{result-3}), the following relation is found,
\begin{equation}
w^{(k)}(\zeta, t) \simeq w^{(i,j)}(\zeta, t) \hspace{1cm} (k=i+j).
\end{equation} From 
this, we conclude that
the insertion of the matrix $B$ on the boundary 
in the two-matrix model corresponds to 
the insertion of one of the matrices $B$ and $C$ in the three-matrix model.
That is, 
there is no essential difference between the insersions of  
the matrices $B$ and $C$,
as far as we consider a finite number of them.
This result shows us that 
some of the local information of the mixed boundary
states cannot be observed in the continuum limit. Therefore,
we cannot distinguish
what kinds of other spin states are inserted.

We know that this type of three-matrix model contains 
not only the $(4, 5)$ model but also
the $(2,7)$ and $(3,8)$ models \cite{Kunitomo}.
This may be related to the fact that the Schwinger-Dyson equation
is of sixth order instead of fourth.  
This is under investigation \cite{AIT}.


We would like to express our gratitude to Prof. M. Ninomiya and 
Prof. Y. Matsuo for warmful encouragemants.
We are grateful to Prof. H. Kunitomo, Prof. M. Fukuma and Dr. F. Sugino
for useful discussions.
Thanks are also due to Prof. M. Ninomiya and M. Fukuma for
careful readings of the manuscript.
This work is supported in part by the Grant-in-Aid for Scientific
Research (2690,5108) from the Ministry of Education, Science and Culture.


\end{document}